\documentclass[12pt,titlepage]{article}
\usepackage{amsfonts}
\usepackage{amsmath}
\usepackage{amssymb}
\usepackage[dvips]{graphicx}

\begin{document}

\title{Scattering and bound states of spin-0 particles in a nonminimal
vector double-step potential }
\date{}
\author{L. P. de Oliveira and A. S. de Castro\thanks{%
castro@pq.cnpq.br.} \\
\\
UNESP - Campus de Guaratinguet\'{a}\\
Departamento de F\'{\i}sica e Qu\'{\i}mica\\
12516-410 Guaratinguet\'{a} SP - Brazil}
\date{ }
\maketitle

\begin{abstract}
The problem of spin-0 particles subject to a nonminimal vector double-step
potential is explored in the context of the Duffin-Kemmer-Petiau theory.
Surprisingly, one can never have an incident wave totally reflected and the
transmission amplitude has complex poles corresponding to bound states. The
interesting special case of bosons embedded in a sign potential with its
unique bound-state solution is analyzed as a limiting case. \newline
\end{abstract}

\section{Introduction}

The first-order Duffin-Kemmer-Petiau (DKP) equation \cite{pet}-\cite{kem} is
often seen as an alternative and rather unusual form for describing spin-0
and spin-1 particles. Although the second-order and the DKP formalisms are
equivalent in the case of minimally coupled vector interactions \cite{mr}-%
\cite{lun}, the last formalism opens news horizons as far as it allows the
inclusion of other kinds of couplings in a straightforward way \cite{gue}-%
\cite{vij}. The nonminimal vector interaction, for instance, refers to a
kind of charge-conjugate invariant coupling which behaves like a vector
under a Lorentz transformation. The invariance of the nonminimal vector
potential under charge conjugation means that it does not couple to the
charge of the boson and so it does not distinguish particles from
antiparticles. Hence, whether one considers spin-0 or spin-1 bosons, this
sort of interaction can not exhibit Klein's paradox \cite{ccc3}. Nonminimal
vector interactions, added by other kinds of Lorentz structures, have
already been used successfully in a phenomenological context for describing
the scattering of mesons by nuclei \cite{cla1}-\cite{cla2}. Nonminimal
vector couplings with diverse functional forms for the potential functions
have been explored in the literature \cite{ccc3}, \cite{Ait}-\cite{asc2}.

In the present work some aspects of the stationary states of spin-0 bosons
in a double-step potential with a nonminimal vector coupling are analyzed.
Scattering states are analyzed and an oscillatory transmission coefficient
is found. An interesting point is that the transmission coefficient never
vanishes, regardless the size of the potential barrier. In sharp contrast
with a nonrelativistic scheme, the transmission amplitude exhibits complex
poles corresponding to bound-state solutions for a potential of sufficient
intensity. The eigenenergies for bound states are solutions of
transcendental equations classified as eigenvalues of the parity operator.
Those intriguing results are interpreted in terms of solutions of an
effective Schr\"{o}dinger equation for a finite square well with additional $%
\delta $-functions situated at the borders. The case of a sign potential
(interpreted as a shifted $\delta $-function potential) with its unique
bound-state solution is analyzed as a limiting case of the double-step
potential.

\section{The nonminimal vector double-step potential}

The DKP equation for a free boson is given by \cite{kem} (with units in
which $\hbar =c=1$)%
\begin{equation}
\left( i\beta ^{\mu }\partial _{\mu }-m\right) \psi =0  \label{dkp}
\end{equation}%
\noindent where the matrices $\beta ^{\mu }$\ satisfy the algebra $\beta
^{\mu }\beta ^{\nu }\beta ^{\lambda }+\beta ^{\lambda }\beta ^{\nu }\beta
^{\mu }=g^{\mu \nu }\beta ^{\lambda }+g^{\lambda \nu }\beta ^{\mu }$
\noindent and the metric tensor is $g^{\mu \nu }=\,$diag$\,(1,-1,-1,-1)$.
That algebra generates a set of 126 independent matrices whose irreducible
representations are a trivial representation, a five-dimensional
representation describing the spin-0 particles and a ten-dimensional
representation associated to spin-1 particles. The second-order Klein-Gordon
and Proca equations are obtained when one selects the spin-0 and spin-1
sectors of the DKP theory. A well-known conserved four-current is given by $%
J^{\mu }=\bar{\psi}\beta ^{\mu }\psi $\noindent $/2$ where the adjoint
spinor $\bar{\psi}$ is given by $\bar{\psi}=\psi ^{\dagger }\eta ^{0}$ with $%
\eta ^{0}=2\beta ^{0}\beta ^{0}-1$. The time component of this current is
not positive definite but it may be interpreted as a charge density. Then
the normalization condition $\int d\tau \,J^{0}=\pm 1$ can be expressed as $%
\int d\tau \,\bar{\psi}\beta ^{0}\psi =\pm 2$, where the plus (minus) sign
must be used for a positive (negative) charge.

With the introduction of nonminimal vector interactions, the DKP equation
can be written as \cite{gue}%
\begin{equation}
\left( i\beta ^{\mu }\partial _{\mu }-m-i[P,\beta ^{\mu }]A_{\mu }\right)
\psi =0  \label{dkp2}
\end{equation}%
where $P$ is a projection operator ($P^{2}=P$ and $P^{\dagger }=P$) in such
a way that $\bar{\psi}[P,\beta ^{\mu }]\psi $ behaves like a vector under a
Lorentz transformation as does $\bar{\psi}\beta ^{\mu }\psi $. Once again $%
\partial _{\mu }J^{\mu }=0$ \cite{ccc3}. If the potential is
time-independent one can write $\psi (\vec{r},t)=\phi (\vec{r})\exp (-iEt)$,
where $E$ is the energy of the boson, in such a way that the
time-independent DKP equation becomes%
\begin{equation}
\left[ \beta ^{0}E+i\beta ^{i}\partial _{i}-\left( m+i[P,\beta ^{\mu
}]A_{\mu }\right) \right] \phi =0  \label{DKP10}
\end{equation}%
For the case of spin 0, we use the representation for the $\beta ^{\mu }$\
matrices given by \cite{ned1}%
\begin{equation}
\beta ^{0}=%
\begin{pmatrix}
\theta & \overline{0} \\
\overline{0}^{T} & \mathbf{0}%
\end{pmatrix}%
,\quad \beta ^{i}=%
\begin{pmatrix}
\widetilde{0} & \rho _{i} \\
-\rho _{i}^{T} & \mathbf{0}%
\end{pmatrix}%
,\quad i=1,2,3  \label{rep}
\end{equation}%
\noindent where%
\begin{eqnarray}
\ \theta &=&%
\begin{pmatrix}
0 & 1 \\
1 & 0%
\end{pmatrix}%
,\quad \rho _{1}=%
\begin{pmatrix}
-1 & 0 & 0 \\
0 & 0 & 0%
\end{pmatrix}
\notag \\
&&  \label{rep2} \\
\rho _{2} &=&%
\begin{pmatrix}
0 & -1 & 0 \\
0 & 0 & 0%
\end{pmatrix}%
,\quad \rho _{3}=%
\begin{pmatrix}
0 & 0 & -1 \\
0 & 0 & 0%
\end{pmatrix}
\notag
\end{eqnarray}%
\noindent $\overline{0}$, $\widetilde{0}$ and $\mathbf{0}$ are 2$\times $3, 2%
$\times $2 and 3$\times $3 zero matrices, respectively, while the
superscript T designates matrix transposition. Here the projection operator
can be written as \cite{gue} $P=\left( \beta ^{\mu }\beta _{\mu }-1\right)
/3=\mathrm{diag}\,(1,0,0,0,0)$. In this case $P$ picks out the first
component of the DKP spinor. The five-component spinor can be written as $%
\psi ^{T}=\left( \psi _{1},...,\psi _{5}\right) $ in such a way that the
time-independent DKP equation for a boson constrained to move along the $X$%
-axis, restricting ourselves to potentials depending only on $x$, decomposes
into%
\begin{equation*}
\left( \frac{d^{2}}{dx^{2}}+E^{2}-m^{2}+A_{0}^{2}-A_{1}^{2}+\frac{dA_{1}}{dx}%
\right) \phi _{1}=0
\end{equation*}%
\begin{equation}
\phi _{2}=\frac{1}{m}\left( E+iA_{0}\right) \,\phi _{1}  \label{dkp4}
\end{equation}%
\begin{equation*}
\phi _{3}=\frac{i}{m}\left( \frac{d}{dx}+A_{1}\right) \phi _{1},\quad \phi
_{4}=\phi _{5}=0
\end{equation*}%
Meanwhile,
\begin{equation}
J^{0}=\frac{E}{m}\,|\phi _{1}|^{2},\quad J^{1}=\frac{1}{m}\text{Im}\left(
\phi _{1}^{\ast }\,\frac{d\phi _{1}}{dx}\right)  \label{corrente4}
\end{equation}%
Given that the interaction potentials satisfy certain conditions, we have a
well-defined Sturm-Liouville problem for determining the possible discrete
or continuous eigenvalues of the system. We also note that there is only one
independent component \ of the DKP spinor for the spin-0 sector. It is not
hard to see that the spectrum is symmetrical about $E=0$, as it should be
since $A_{\mu }$ does not distinguish particles from antiparticles. Note
also that for ensuring the covariance of the DKP theory under the parity
operation one must have $A_{0}\left( -x\right) =+A_{0}\left( x\right) $ and $%
A_{1}\left( -x\right) =-A_{1}\left( x\right) $. It follows that the parity
of $\phi _{3}$ is opposite to that one of $\phi _{1}$ and $\phi _{2}$ in
such a way that the DKP spinor has a definite parity. Furthermore, the
change $A_{\mu }\rightarrow A_{\mu }+constant$ drastically changes the
spectrum.

Let us focus our attention on the space component of a nonminimal potential
with $A_{1}\left( -x\right) =-A_{1}\left( x\right) $. We consider the
double-step potential
\begin{equation}
A_{1}\left( x\right) =V_{0}\left[ \theta \left( x-a\right) -\theta \left(
-x-a\right) \right]   \label{sc1a}
\end{equation}%
with $V_{0}$ and $a$ defined to be real numbers ($a>0$) and $\theta \left(
x\right) $ is the Heaviside step function. It is of interest to note that in
the limit $a\rightarrow 0$, the double-step function reduces to $V_{0}\,$%
\textrm{sgn}$\left( x\right) $, where $\,$\textrm{sgn}$\left( x\right) =x/|x|
$. Our problem is to solve (\ref{dkp4}) for $\phi _{1}$ and to determine the
allowed energies. In this case the first line of (\ref{dkp4}) can be written
as%
\begin{equation*}
\frac{d^{2}\phi _{1}}{dx^{2}}+\left\{ E^{2}-m^{2}+V_{0}\left[ \delta \left(
x-a\right) +\delta \left( x+a\right) \right] \right.
\end{equation*}%
\begin{equation}
\qquad \qquad \qquad \qquad \qquad \qquad \qquad \left. -V_{0}^{2}\left[
\theta \left( x-a\right) +\theta \left( -x-a\right) \right] \right\} \phi
_{1}=0  \label{eq1}
\end{equation}%
where $\delta \left( x\right) =d\theta \left( x\right) /dx$ is the Dirac
delta function. We turn our attention to scattering states so that the
solutions describing spinless bosons coming from the left can be written as
\begin{equation}
\phi _{1}(x)=\left\{
\begin{array}{cc}
Ae^{+i\eta \frac{x}{a}}+Be^{-i\eta \frac{x}{a}} & \text{\textrm{for }}x<-a
\\
&  \\
Ce^{+i\xi \frac{x}{a}}+De^{-i\xi \frac{x}{a}} & \text{\textrm{for }}|x|<a \\
&  \\
Fe^{+i\eta \frac{x}{a}} & \text{\textrm{for }}x>a%
\end{array}%
\right.   \label{eq50}
\end{equation}%
\noindent where%
\begin{equation}
\xi =a\sqrt{E^{2}-m^{2}},\quad \eta =\sqrt{\xi ^{2}-\mathcal{\upsilon }^{2}}%
,\quad \mathcal{\upsilon }=aV_{0}
\end{equation}%
\noindent Then, $\phi _{1}$ describes an incident wave moving to the right ($%
\eta $ is a real number) and a reflected wave moving to the left with%
\begin{equation}
J^{1}\left( x<-a\right) =\frac{\eta }{am}\left( |A|^{2}-|B|^{2}\right)
\end{equation}%
and a transmitted wave moving to the right with%
\begin{equation}
J^{1}\left( x>a\right) =\frac{\eta }{am}|F|^{2}
\end{equation}%
We demand that $\phi _{1}$ be continuous at $x=\pm a$, i. e.%
\begin{equation}
\lim_{\varepsilon \rightarrow 0}\left. \phi _{1}\right\vert _{x=\pm
a-\varepsilon }^{x=\pm a+\varepsilon }=0  \label{cont}
\end{equation}%
Otherwise, (\ref{eq1}) would contain the derivative of a $\delta $-function.
Effects due to the potential on $d\phi _{1}/dx$ in the neighbourhood of $\
x=\pm a$ can be evaluated by integrating (\ref{eq1}) from $\pm a-\varepsilon
$ to $\pm a+\varepsilon $ and taking the limit $\varepsilon \rightarrow 0$.
The connection formula between $d\phi _{1}/dx$ at the right and $d\phi
_{1}/dx$ at the left can be summarized as%
\begin{equation}
\lim_{\varepsilon \rightarrow 0}\left. \frac{d\phi _{1}}{dx}\right\vert
_{x=\pm a-\varepsilon }^{x=\pm a+\varepsilon }=-\frac{\mathcal{\upsilon }}{a}%
\,\phi _{1}(\pm a)  \label{discont}
\end{equation}%
With $\phi _{1}$ given by (\ref{eq50}), conditions (\ref{cont}) and (\ref%
{discont}) imply that
\begin{eqnarray}
e^{i\eta }F &=&e^{i\xi }C+e^{-i\xi }D  \notag \\
&&  \notag \\
e^{-i\eta }A+e^{i\eta }B &=&e^{-i\xi }C+e^{i\xi }D  \notag \\
&& \\
e^{i\eta }F\left( \eta -i\mathcal{\upsilon }\right)  &=&\xi \left( e^{i\xi
}C-e^{-i\xi }D\right)   \notag \\
&&  \notag \\
\xi \left( e^{-i\xi }C-e^{i\xi }D\right)  &=&\left( \eta +i\mathcal{\upsilon
}\right) e^{-i\eta }A-\left( \eta -i\mathcal{\upsilon }\right) e^{i\eta }B
\notag
\end{eqnarray}%
Omitting the algebraic details, we state the solution for the relative
amplitudes%
\begin{eqnarray}
\frac{B}{A} &=&\frac{2ie^{-2i\eta }\xi \mathcal{\upsilon }\cos 2\xi }{d}
\notag \\
&&  \notag \\
\frac{C}{A} &=&\frac{e^{-i\xi }e^{-i\eta }\eta \left( \xi +\eta -i\mathcal{%
\upsilon }\right) }{d}  \notag \\
&&  \label{amp} \\
\frac{D}{A} &=&\frac{e^{i\xi }e^{-i\eta }\eta \left( \xi -\eta +i\mathcal{%
\upsilon }\right) }{d}  \notag \\
&&  \notag \\
\frac{F}{A} &=&\frac{2e^{-2i\eta }\xi \eta }{d}  \notag
\end{eqnarray}%
where we have defined%
\begin{equation}
d\equiv 2\left( \eta -i\mathcal{\upsilon }\right) \left( \xi \cos 2\xi
-i\eta \sin 2\xi \right)
\end{equation}

In order to determinate the reflection and transmission coefficients we use
the charge current densities $J^{1}\left( x<-a\right) $ and $J^{1}\left(
x>a\right) $. The $x$-independent current density allow us to define the
reflection and transmission coefficients as%
\begin{equation}
R=\left\vert \frac{B}{A}\right\vert ^{2},\quad T=\left\vert \frac{F}{A}%
\right\vert ^{2}
\end{equation}%
with $R+T=1$. The last relative amplitude in (\ref{amp}) allow us to write
the transmission coefficient as%
\begin{equation}
T=\left[ 1+\left( \frac{\mathcal{\upsilon }}{\eta }\cos 2\xi \right) ^{2}%
\right] ^{-1}
\end{equation}%
regardless of the sign of $\mathcal{\upsilon }$. Notice that $T\rightarrow 1$
as $\eta \rightarrow \infty $ and that there is a resonance transmission ($%
T=1$) whenever $\xi =\left( 2n+1\right) \pi /4$ with $n=0,1,2,\ldots $

The possibility of bound states requires a solution given by (\ref{eq50})
with $\eta =i|\eta |$ ($\xi <|\mathcal{\upsilon }|$) and $A=0$ in order to
obtain a square-integrable $\phi _{1}$. Therefore, if one considers the
transmission amplitude as a function of the complex variable $\eta $ one
sees that for $\eta $ real and positive one obtains the scattering states
whereas the bound states would be obtained by the poles lying along the
positive imaginary axis of the complex $\eta $-plane. Setting $|\eta |=\sqrt{%
\mathcal{\upsilon }^{2}-\xi ^{2}}$ and expanding $F/A$ in a power series in $%
\xi $ about $\xi =0$, we obtain%
\begin{equation}
\frac{F}{A}=\frac{e^{2|\mathcal{\upsilon }|}|\mathcal{\upsilon }|}{\left( |%
\mathcal{\upsilon }|-\mathcal{\upsilon }\right) \left( 1+2|\mathcal{\upsilon
}|\right) }+O\left( \xi \right)
\end{equation}%
where $O\left( \xi \right) $ denotes higher-order terms. Thus, $\xi =0$ is a
pole only for $\mathcal{\upsilon }>0$. The other poles are solutions of the
transcendental equation%
\begin{equation}
\tan 2\xi =-\frac{\xi }{|\eta |},\quad \xi \neq 0
\end{equation}%
With the amplitudes given by (\ref{amp}) one obtains the ratios
\begin{eqnarray}
\frac{C}{D} &=&\frac{\xi +\eta -i\mathcal{\upsilon }}{\xi -\eta +i\mathcal{%
\upsilon }}e^{-2i\xi }  \notag \\
&& \\
\frac{B}{F} &=&i\frac{\mathcal{\upsilon }}{\eta }\cos 2\xi  \notag
\end{eqnarray}%
so that one can write
\begin{eqnarray}
\frac{C}{D} &=&e^{-i\left( 2\xi +\arctan \frac{\xi }{|\eta |}\right) }
\notag \\
&&  \label{Amp} \\
\frac{B}{F} &=&\frac{\mathcal{\upsilon }}{|\eta |}\cos 2\xi  \notag
\end{eqnarray}%
for $\eta =i|\eta |$. It is true that the first line of (\ref{Amp})
furnishes $|C|=|D|$. It has to be so since the charge current density $J^{1}$
vanishes in the region $|x|>a$ whereas in the region $|x|<a$ it takes the
form%
\begin{equation}
J^{1}=\left\{
\begin{array}{cc}
\frac{\xi }{am}\left( |C|^{2}-|D|^{2}\right) & \text{\textrm{for }}\xi \text{
\textrm{real}} \\
&  \\
\frac{2|\xi |}{am}\text{\thinspace \textrm{Im}}\left( C^{\ast }D\right) &
\text{\textrm{for }}\xi \text{ \textrm{imaginary}}%
\end{array}%
\right.
\end{equation}%
Hence, one concludes that bound states are only possible if $\xi =0$ or $%
|C|=|D|$. Up to this point symmetry arguments has not come into the story at
all. Since $A_{1}(x)$ is antisymmetric with respect to $x$, it follows that $%
\phi _{1}$ can be either even or odd. Hence,
\begin{equation}
\frac{C}{D}=\frac{B}{F}=\pm 1  \label{ratio}
\end{equation}%
so that%
\begin{equation}
\phi _{1}\left( x\right) =\left\{
\begin{array}{cc}
+Fe^{+|\eta |\frac{x}{a}} & \text{\textrm{for }}x<-a \\
&  \\
2C\cos \left( \xi \frac{x}{a}\right) & \text{\textrm{for }}|x|<a \\
&  \\
+Fe^{-|\eta |\frac{x}{a}} & \text{\textrm{for }}x>+a%
\end{array}%
\right.
\end{equation}%
for $\phi _{1}\left( -x\right) =+\phi _{1}\left( x\right) $, and%
\begin{equation}
\phi _{1}\left( x\right) =\left\{
\begin{array}{cc}
-Fe^{+|\eta |\frac{x}{a}} & \text{\textrm{for }}x<-a \\
&  \\
2iC\sin \left( \xi \frac{x}{a}\right) & \text{\textrm{for }}|x|<a \\
&  \\
+Fe^{-|\eta |\frac{x}{a}} & \text{\textrm{for }}x>+a%
\end{array}%
\right.
\end{equation}%
for $\phi _{1}\left( -x\right) =-\phi _{1}\left( x\right) $. The condition $%
C/D=\pm 1$ demands%
\begin{equation}
\tan 2\xi =-\frac{\xi }{|\eta |},\quad \xi \geq 0
\end{equation}%
Using the identity $\tan 2z=2\tan z/\left( 1-\tan ^{2}z\right) $, one can
write this last relation as%
\begin{equation}
\xi \tan \xi =|\eta |-\lambda |\mathcal{\upsilon }|,\quad \lambda =\pm 1
\label{bs}
\end{equation}%
with the proviso that the root $\xi =0$ is valid only for $\lambda =+1$
(from (\ref{Amp}) one sees that $C/D=+1$ for $\xi =0$). In addition, by
virtue of the identity $\cos 2z=\left( 1-\tan ^{2}z\right) /\left( 1+\tan
^{2}z\right) $, one may readily check that (\ref{bs}) implies into%
\begin{equation}
\frac{B}{F}=\lambda \,\mathrm{sgn}\left( \mathcal{\upsilon }\right)
\end{equation}%
Now we see that $\lambda \,$\textrm{sgn}$\left( \mathcal{\upsilon }\right) $
is the parity eigenvalue and that $\xi =0$ furnishes a legitimate even
bound-state solution for $\mathcal{\upsilon }>0$ (whatever the intensity of $%
\upsilon $, the solution $\xi =0$ makes $J^{0}$ independent of $x$ for $%
|x|<a $). It is instructive to note that, except for $\xi =0$, the equation
for even-parity solutions is mapped into that one for odd-parity solutions
under the change of \ $\mathcal{\upsilon }$\ by $-\mathcal{\upsilon }$, and
vice versa. In addition, because $|\eta |=\sqrt{\mathcal{\upsilon }^{2}-\xi
^{2}}$, (\ref{bs}) can also be written as%
\begin{equation}
-\xi \cot \xi =|\eta |+\lambda |\mathcal{\upsilon }|,\quad \lambda =\pm 1
\end{equation}%
Based on $\tanh z=-i\tan iz$, equation (\ref{bs}) is transformed into%
\begin{equation}
-|\xi |\tanh |\xi |=|\eta |-\lambda |\mathcal{\upsilon }|,\quad \lambda =\pm
1  \label{b2}
\end{equation}%
for $\xi =i|\xi |$ ($|\eta |=\sqrt{\mathcal{\upsilon }^{2}+|\xi |^{2}}$).
Equation (\ref{b2}) is the quantization condition corresponding to $%
\varepsilon <0$. The solutions of these transcendental equations are to be
found from numerical or graphical methods.

Tackling (\ref{b2}) first, one sees that except for $\xi =0$ for $\mathcal{%
\upsilon }>0$ and even $\phi _{1}$, or $\mathcal{\upsilon }<0$ and odd $\phi
_{1}$, the left-hand side of (\ref{b2}) is always negative whereas its
right-hand side is always positive. Therefore, equations (\ref{b2}) furnish
no solutions, except $\xi =0$ for even\textrm{\ }$\phi _{1}$ with $\mathcal{%
\upsilon }>0$.

The graphical method for $\xi \in \,$$\mathbb{R}$ is illustrated in Figure
1. The solutions for bound states are given by the intersection of the curve
represented by $\xi \tan \xi $ with the curves represented by $|\eta
|-\lambda |\mathcal{\upsilon }|$. We can see immediately that, except for $%
\xi =0$ for $\lambda =+1$, it needs critical values, corresponding to $|%
\mathcal{\upsilon }_{\text{c}}|=-\arctan \lambda $, for obtaining bound
states. If $\mathcal{\upsilon }$ is larger than the critical values there
will be a finite sequence of bound states with alternating parities. The
ground-state solution will correspond to an even solution. As commented
before, the odd-parity solution with $\lambda =+1$ is spurious. When $%
\upsilon $ approaches infinity the intersections will occur at the
asymptotes of $\xi \tan \xi $ (except for $\xi =0$) so that the solutions
will be given by
\begin{equation}
\xi _{n}^{\left( \infty \right) }=\left( 2n+2\right) \frac{\pi }{4},\qquad
n=0,1,2,\ldots  \label{eq110}
\end{equation}%
\noindent Then, if
\begin{equation}
\frac{\pi }{4}\leq |\mathcal{\upsilon }|<\left( 2N+1\right) \frac{\pi }{4}
\label{eq110a}
\end{equation}%
\noindent there will be $N$ bound states, except for $\xi =0$ for $\mathcal{%
\upsilon }>0$, with eigenvalues given by
\begin{equation}
\left( 2N-1\right) \frac{\pi }{4}\leq \xi _{N}<N\frac{\pi }{2}
\label{eq110b}
\end{equation}

Finally, in the limit $a\rightarrow 0$ ($\xi \rightarrow 0$) one has%
\begin{equation}
T\underset{a\rightarrow 0}{\longrightarrow }\left( 1+\frac{V_{0}^{2}}{%
E^{2}-m^{2}-V_{0}^{2}}\right) ^{-1}
\end{equation}%
and the transmission amplitude has one and only one pole (at $|E|=m$ for $%
V_{0}>0$) corresponding to an even bound-state solution with
\begin{equation}
\phi _{1}\left( x\right) \underset{a\rightarrow 0}{\longrightarrow }%
Fe^{-V_{0}|x|}
\end{equation}

\section{Concluding remarks}

The stationary states of spinless bosons interacting via nonminimal vector
coupling was investigated by a technique which maps the DKP equation into a
Sturm-Liouville problem for the first component of the DKP spinor.
Scattering states in a double-step potential were analyzed and an
oscillatory transmission coefficient was found. An interesting feature of
the scattering is that the transmission coefficient never vanishes, no
matter how large $V_{0}$ may be. It was shown that, for a potential of
sufficient intensity, the transmission amplitude exhibits complex poles
corresponding to bound-state solutions. The eigenenergies for bound states
are solutions of transcendental equations classified as eigenvalues of the
parity operator. The case of a sign potential was analyzed by a limiting
process. In that last case we obtained a non-oscillatory transmission
coefficient and a unique bound state. As the potential is a double step, or
a sign potential in a limiting case, one should not expect the existence of
bound states, and it follows that such bound states are consequence of the
peculiar coupling in the DKP equation.

For a better understanding of \ those unexpected results, it can be observed
that the first line of (\ref{dkp4}) can also be written as
\begin{equation}
\left[ -\frac{1}{2m}\frac{d^{2}}{dx^{2}}+V_{\mathtt{eff}}\left( x\right) %
\right] \phi _{1}\,=E_{\mathtt{eff}}\phi _{1}  \label{Heff}
\end{equation}%
with%
\begin{equation}
E_{\mathtt{eff}}=\frac{E^{2}-m^{2}}{2m},\qquad V_{\mathtt{eff}}\left(
x\right) =\frac{A_{1}^{2}-A_{0}^{2}-dA_{1}/dx}{2m}  \label{sc2}
\end{equation}%
The set (\ref{Heff})-(\ref{sc2}) plus $\int_{-\infty }^{+\infty }dx\,|\phi
_{1}|^{2}<\infty $ correspond to the nonrelativistic description of a
particle of mass $m$ with energy $E_{\mathtt{eff}}$ subject to a potential $%
V_{\mathtt{eff}}$. As the effective potential has a more complicated
structure, with quadratic plus derivative terms, the success of the strategy
of this mapping depends crucially of the choice for the potential $A_{\mu
}\left( x\right) $. Examination of the double-step potential given by (\ref%
{sc1a}) shows that
\begin{equation}
V_{\mathtt{eff}}\left( x\right) =\frac{V_{0}^{2}}{2m}\left[ \theta \left(
x-a\right) +\theta \left( -x-a\right) \right] -\frac{V_{0}}{2m}\left[ \delta
\left( x-a\right) +\delta \left( x+a\right) \right]  \label{eff}
\end{equation}%
Therefore one has to search for solutions of the Schr\"{o}dinger equation
for a particle under the influence of a finite square well potential with
attractive (repulsive) $\delta $-functions when $V_{0}>0$ ($V_{0}<0$)
situated at the borders. Whether $V_{0}$ is positive or negative, the
effective potential $V_{\mathtt{eff}}\left( x\right) $ has also a form that
would make allowance for bound-state solutions with $E_{\mathtt{eff}%
}<V_{0}^{2}/\left( 2m\right) $, and $E_{\mathtt{eff}}>0$ if $V_{0}<0$. For $%
a\rightarrow 0$, the case of a sign potential, the effective potential
becomes the shifted $\delta $-function potential:%
\begin{equation}
V_{\mathtt{eff}}\left( x\right) \underset{a\rightarrow 0}{\longrightarrow }%
\frac{V_{0}^{2}}{2m}-\frac{V_{0}}{m}\delta \left( x\right)
\end{equation}%
which leads to a non-oscillatory transmission coefficient independent of the
sign of $V_{0}$, and for $V_{0}>0$ to exactly one bound-state solution, that
one with a vanishing effective energy ($|E|=m$) independent of the size of $%
V_{0}$.

\bigskip \bigskip \bigskip \bigskip \bigskip \bigskip

\noindent {\textbf{Acknowledgments}}

This work was supported in part by means of funds provided by Coordena\c{c}%
\~{a}o de Aperfei\c{c}oamento de Pessoal de N\'{\i}vel Superior\ (CAPES) and
Conselho Nacional de Desenvolvimento Cient\'{\i}fico e Tecnol\'{o}gico
(CNPq).

\newpage

\begin{figure}[th]
\begin{center}
\includegraphics[width=9cm, angle=0]{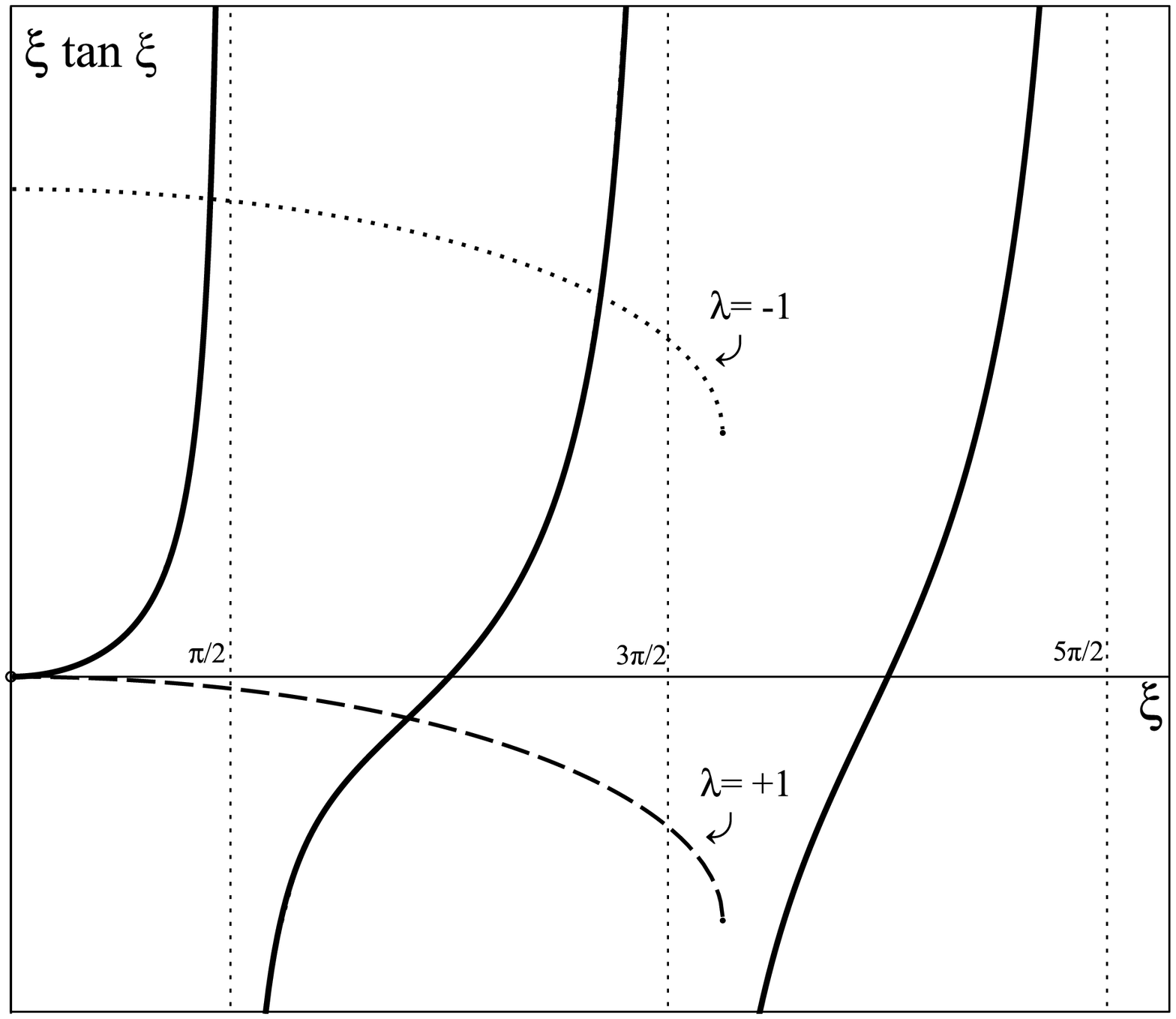}
\end{center}
\par
\vspace*{-0.1cm}
\caption{Graphical solution of $\protect\xi \tan \protect\xi =|\protect\eta %
|-\protect\lambda |\mathcal{\protect\upsilon }|$ for the case $|\mathcal{%
\protect\upsilon }|=5$. The solid lines stand for $\protect\xi \tan \protect%
\xi$, the dotted line for $|\protect\eta |+ |\mathcal{\protect\upsilon }|$
and the dashed line for $|\protect\eta |-|\mathcal{\protect\upsilon }|$. The
root $\protect\xi =0$ is valid only for $\mathcal{\protect\upsilon }>0$.}
\label{Fig1}
\end{figure}

\end{document}